\documentclass[preprint,12pt]{elsarticle}

\usepackage{amssymb}
\usepackage{amsmath}

\usepackage{booktabs}
\usepackage{rotating}
\usepackage{rotfloat}
\usepackage{csquotes}
\usepackage{xcolor}
\newenvironment{revision_1}{\color{black}}{}

\usepackage[normalem]{ulem} 

\usepackage{tabularx}
\newlength{\mylandscapewidth}
\journal{Elsevier}

\begin{document}

\begin{frontmatter}



\title{\textbf{Next-Generation Segmentation in Radiotherapy: 
\begin{revision_1}
A Narrative Review from Atlas-Based Methods to Agentic 
Vision--Language Models
\end{revision_1}
}}


\author[1]{Josef A. Buchner\corref{cor1}}
\cortext[cor1]{Corresponding author}
\ead{j.buchner@tum.de}
\author[1,2]{Richard Osuala}
\author[1,3]{Ayhan Can Erdur}
\author[1]{Melissa C. Olufs}
\author[1,4]{Stephanie E. Combs}
\author[1,4]{Jan C. Peeken}

\affiliation[1]{organization={Department of Radiation Oncology, TUM Klinikum rechts der Isar, Technical University of Munich}, city={Munich}, country= {Germany}}
\affiliation[2]{organization={Artificial Intelligence in Medicine Lab (BCN-AIM), Departament de Matemàtiques i Informàtica, Universitat de Barcelona}, city={Barcelona}, country={Spain}}
\affiliation[3]{organization={Chair for AI in Healthcare and Medicine, TUM Klinikum rechts der Isar, Technical University of Munich}, city={Munich}, country={Germany}}
\affiliation[4]{organization={Partner Site Munich, German Consortium for Translational Cancer Research (DKTK)}, city={Munich}, country={Germany}}

\begin{abstract}
Precise segmentation of target volumes and organs at risk is an essential prerequisite for modern, high-precision radiation therapy.
However, this task has traditionally been time-consuming and subject to high interobserver variability.
Motivated by these limitations, more advanced and novel approaches to automating this task have been developed over the past two decades.
This narrative review highlights representative milestones, with a focus on the most recent developments.
These approaches include atlas-based label propagation, convolutional architectures, promptable foundation models, and, most recently, \begin{revision_1}vision--language models\end{revision_1} (VLMs), which integrate clinical text with volumetric segmentation.
Each methodological generation has addressed limitations of its predecessor while introducing new ones. 
A substantial gap remains between research progress and clinical deployment, as routine workflows continue to be dominated by atlas-based and narrow task-specific deep-learning tools\begin{revision_1}, which remain strong baselines in geometric overlap metrics but are highly rigid in terms of application, versatility, portability, and interpretability.\end{revision_1}
As \begin{revision_1}contouring\end{revision_1} systems \begin{revision_1}further\end{revision_1} grow in complexity, conventional geometric overlap metrics become insufficient, motivating evaluation paradigms that incorporate surface-based, dosimetric, and prompt-adherence assessment.
Emerging agentic architectures that combine large language models, retrieval-augmented generation, and vision--language segmentation modules are leading towards \begin{revision_1} more flexible,\end{revision_1} guideline-aware, interactive contouring workflows. 
These systems could automate flexible \begin{revision_1} and interpretable \end{revision_1} knowledge-based contouring and actively support the clinical reasoning underlying target volume definition.
\end{abstract}




\end{frontmatter}

\section*{Funding Statement}
This research was supported by the European Research Council (ERC) ERC-2025-STG Horizon Grant (101219993) and Else Kr\"oner-Fresenius Foundation (2023\_EKES.01).

\section*{Conflict of Interest Statement}
\textbf{JAB:} Funding: ERC; 
\textbf{RO:} \begin{revision_1} Funding: ERC; \end{revision_1}
\textbf{ACE:} none; 
\textbf{MCO:} Funding: ERC; 
\textbf{SEC:} Grants: DFG Emmy Noether Programme, BMBF Grant; Consulting Fees, Payment or honoraria: Icotec AG, Roche, BMS, Brainlab, AstraZeneca, Dr. Sennewald Medizintechnik, Daiichi Sankyo, Accuray, Elekta, Medac, Janssen Cilag, Novocure, CureVac, Merck, Nexstim, HMG Systems Engineering, PGXperts GmbH, Zeiss Meditec AG, Servier, medupdate GmbH, OncoMAGNETx; Advisory board: CureVac DSMB Member, ALA-RDT in GBM DSMB Member; Leadership: DEGRO and NOA Board Member, Brainlab Member Supervisory Board; all not related not current manuscript; 
\textbf{JCP:} Funding: ERC, Else Kr\"oner-Fresenius Foundation; Payment or honoraria: AstraZeneca, Brainlab; Stock: Mevidence GmbH



\section{Introduction}
Radiation therapy is a cornerstone of modern cancer treatment, with approximately 50\% of cancer patients receiving it at some point during their illness \cite{delaney2005role, zhu2024global}.
Recent advances in radiation oncology, including intensity-modulated radiation therapy and image-guided radiation therapy, now enable dose delivery with millimeter precision \cite{gregoire2020image}, thereby reducing dose to organs at risk (OARs) and treatment-related toxicity \cite{nutting2011parotid}.
Such conformal irradiation, however, demands accurate delineation of both the clinical target volumes (CTVs) and OARs \cite{weiss2003impact, njeh2008tumor}.

Historically, the manual creation of these segmentations has been a time-consuming and labor-intensive component of the clinical workflow \cite{harari2010emphasizing}.
Moreover, manual delineation is inherently prone to interobserver variability, arising from ambiguous anatomical boundaries and differences in individual contouring style \cite{weiss2003impact, njeh2008tumor}.

\begin{revision_1}
    It is important to distinguish the three principal segmentation tasks in radiotherapy, as they differ fundamentally in complexity and in the type of information required.
    OAR segmentation is primarily anatomy-driven, as the structures are visible on imaging and their boundaries are largely defined by image contrast.
    Gross tumor volume (GTV) delineation also targets image-visible structures, but the boundaries are often ambiguous, patient-specific, and dependent on imaging modality and quality.
\end{revision_1}

In contrast to \begin{revision_1}both these tasks\end{revision_1}, CTVs are not directly visible on the imaging modalities.
The extent of these regions is therefore not directly observable but must instead be inferred from tumor biology, lymphatic drainage patterns, and individual patient characteristics such as tumor stage, histology, and prior surgical interventions \cite{burnet2004defining}.
This dependence on clinical reasoning rather than observable imaging features alone adds a further layer of uncertainty on top of the variability already present in GTV delineation, making CTV definition considerably more susceptible to interobserver variability than the delineation of largely image-defined OARs \cite{weiss2003impact, njeh2008tumor}, with national review projects reporting marked disagreement in CTV delineation across radiotherapy centers \cite{van2019interobserver, bollen2023impact}.
To reduce this variability and ensure consistent, evidence-based target definitions, international bodies such as the European Society for Therapeutic Radiology and Oncology (ESTRO) and the Radiation Therapy Oncology Group (RTOG) have developed standardized contouring guidelines. 
Adherence to such guidelines has been shown to substantially reduce interobserver variability \cite{bollen2023impact, beck2022adherence}, which underlines their critical role in clinical practice.
The resulting CTV segmentations form the basis of treatment planning, in which dose distributions are optimized to ensure adequate target coverage and minimize exposure to surrounding healthy tissue, which leads to reduced treatment-related side effects \cite{burnet2004defining}. 
\begin{revision_1}
This three-level distinction is maintained throughout the review, as each methodological generation addresses these tasks to different degrees.
\end{revision_1}

This review focuses on how segmentation in radiotherapy has evolved from atlas-based label propagation to task-specific deep learning and, more recently, to promptable and multimodal models that aim to incorporate user intent.

\begin{revision_1}
    It is intended as complementary to the NRG Oncology assessment by Rong \textit{et al.} \cite{rong2024nrg}, which provides a comprehensive evaluation of commercially available deep learning auto-segmentation tools with a focus on clinical implementation, quality assurance, and standardization, and to a recent review of CNN-based auto-segmentation for OAR, GTV, and CTV delineation \cite{erdur2025deep}.
    The present review instead traces the methodological trajectory from current clinical tools toward emerging research paradigms, with particular attention to the transition from purely vision-based to language-conditioned and agentic architectures.
    Beyond this overview, it makes two radiotherapy-specific contributions. 
    First, we outline an agentic, retrieval-augmented contouring workflow in which contouring guidelines are treated as an external, updatable knowledge base rather than being implicitly encoded in model weights. 
    This addresses the clinical reality that guidelines evolve and, to our knowledge, distinguishes our review from prior agentic radiotherapy work, which has focused on treatment planning rather than target definition. 
    Second, we propose an evaluation paradigm for language-conditioned segmentation that extends beyond geometric overlap, comprising counterfactual prompt testing, language ablation, and semantic invariance testing, complemented by effort-aware, dosimetric, and outcome-linked endpoints.
\end{revision_1}

\begin{figure}[htbp]
  \centering
  \includegraphics[width=\textwidth]{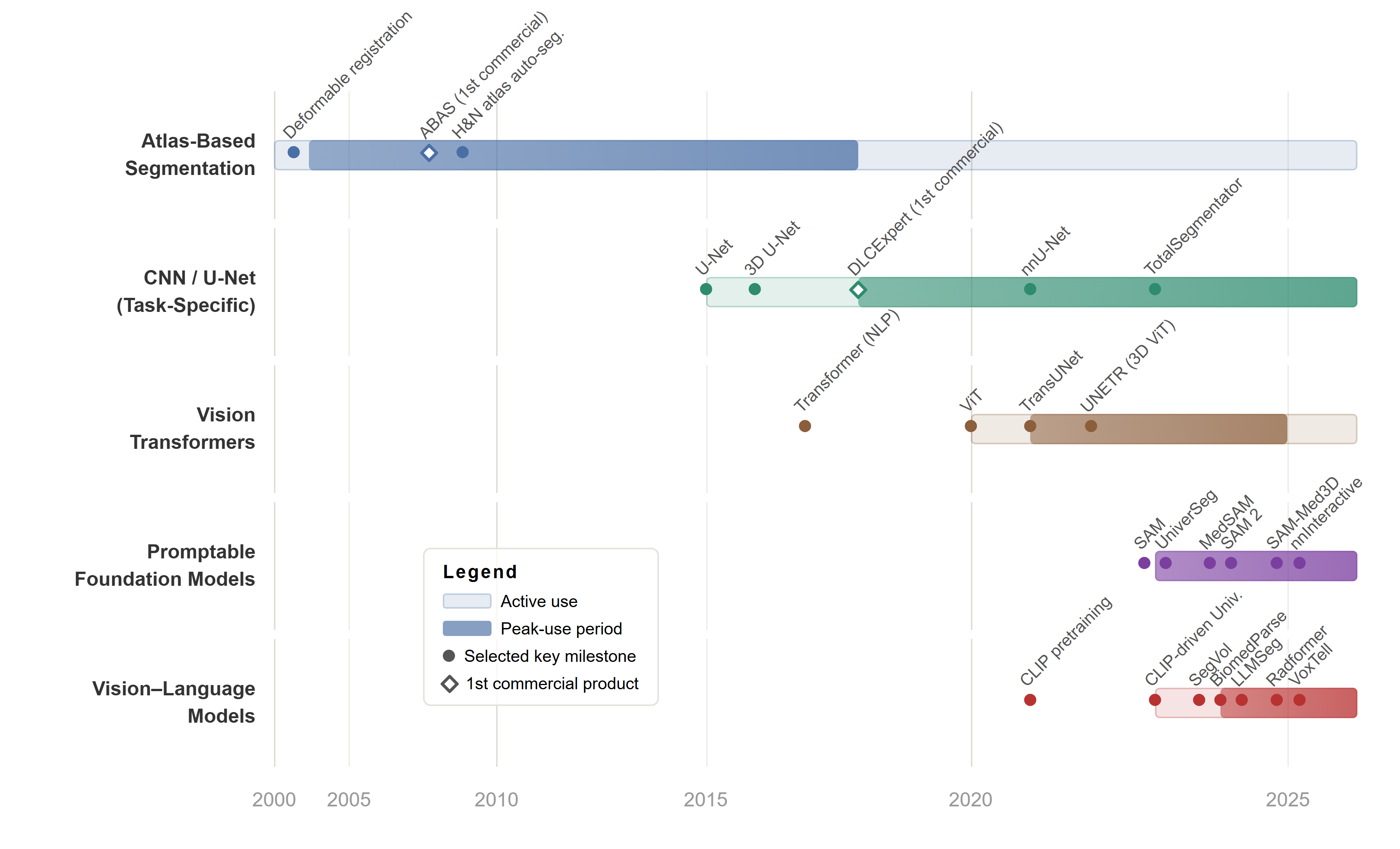}
  \caption[Parallel Evolution of Segmentation Methodologies in Radiation Therapy.]{\textbf{Parallel Evolution of Segmentation Methodologies in Radiation Therapy.} 
  Year ranges are approximate and reflect first peer-reviewed publications. 
  Clinical adoption commonly lags several years behind research publication. 
  Selected commercial milestones are marked separately. 
  NLP = Natural Language Processing, SAM = Segment Anything Model \cite{kirillov2023segment}, ViT = Vision Transformer \cite{dosovitskiy2020image}.}
  \label{fig:timeline}
\end{figure}

Rather than aiming for exhaustive coverage, the review highlights representative milestones that illustrate this transition (Figure~\ref{fig:timeline}); the main characteristics of representative approaches are compared in Table~\ref{tab:comparison}, and the literature selection is described in Section~\ref{sec:search}.

\begin{sidewaystable}[p]
\centering
\caption{Comparison of representative segmentation approaches discussed in this review.
SAM/SAM\,2 (natural-image models) \cite{kirillov2023segment, ravi2024sam}, TransUNet \cite{chen2021transunet}, UNETR \cite{hatamizadeh2022unetr}, MedSAM\,2 \cite{zhu2024medical}, and BiomedParse-V \cite{zhao2025biomedparse} are discussed in the text but omitted here for brevity.}
\label{tab:comparison}
\scriptsize
\setlength{\tabcolsep}{3pt}
\renewcommand{\arraystretch}{1.35}
\begin{tabular*}{\textheight}{@{\extracolsep{\fill}} l c p{1.8cm} p{2.2cm} p{2.6cm} p{3.0cm} p{2.2cm} p{3.2cm} @{}}
\toprule
\textbf{Model} &
\textbf{Year} &
\textbf{Modality} &
\textbf{Input Dim.} &
\textbf{Interaction} &
\textbf{Role of Language} &
\textbf{Label Space} &
\textbf{Main Limitation} \\
\midrule
 
\multicolumn{8}{l}{\textit{Atlas-Based Methods}} \\[2pt]
 
Multi-Atlas Segm. \cite{iglesias2015multi}
  & 2006
  & CT, MRI
  & Native 3D
  & Automatic
  & ---
  & Fixed (atlas labels)
  & Registration fails with large anatomical deviations\\
 
\midrule
\multicolumn{8}{l}{\textit{CNN / U-Net (Task-Specific)}} \\[2pt]
 
nnU-Net \cite{isensee2021nnu}
  & 2021
  & Any
  & Auto-config.\ (2D, 3D, cascade)
  & Automatic
  & ---
  & Fixed (per task)
  & Task-specific training; limited long-range context \\
 
TotalSegmentator \cite{wasserthal2023totalsegmentator}
  & 2023
  & CT
  & Native 3D
  & Automatic
  & ---
  & Fixed (104 classes)
  & CT only; fixed label set; no CTV support \\
 
\midrule
\multicolumn{8}{l}{\textit{Promptable Foundation Models}} \\[2pt]
 
UniverSeg \cite{butoi2023universeg}
  & 2023
  & Multi-mod.\ medical (16 mod.)
  & 2D ($128^2$)
  & In-context (support set of image--label pairs)
  & ---
  & Defined by support set
  & 2D binary only; quality depends on support set \\
 
MedSAM \cite{ma2024segment}
  & 2024
  & Multi-mod.\ medical
  & 2D (slice-by-slice)
  & Interactive (bound.\ boxes)
  & ---
  & Class-agnostic
  & 2D only; per-slice bounding boxes required \\
 
SAM-Med3D \cite{wang2025sam}
  & 2025
  & Multi-mod.\ medical
  & Native 3D (crop)
  & Interactive (vol.\ points)
  & ---
  & Class-agnostic
  & Small crop size ($128^3$); sliding-window inference \\
 
nnInteractive \cite{isensee2025nninteractive}
  & 2025
  & CT, MRI, PET
  & Native 3D
  & Interactive, iterative (points, boxes, scribbles, lasso)
  & ---
  & User prompt-defined spatial label set
  & No semantic prompting; spatial prompts only \\
 
\midrule
\multicolumn{8}{l}{\textit{Vision--Language Models}} \\[2pt]
 
CLIP-Driven Univ. \cite{liu2023clip}
  & 2023
  & CT
  & Native 3D
  & Automatic
  & Label embedd.\ as semantic regularizer
  & Fixed (CLIP-enhanced)
  & No text prompting; fixed class set at inference \\
 
SegVol \cite{du2024segvol}
  & 2024
  & CT (partial MRI)
  & Native 3D
  & Interactive (points, boxes + text)
  & Semantic prompts via frozen CLIP (category names)
  & Semi-open ({$>$}200 cat.)
  & Frozen CLIP encoder limits medical vocabulary \\
 
BiomedParse \cite{zhao2025foundation}
  & 2025
  & Multi-mod.\ (9 mod.)
  & 2D only
  & Automatic (text-prompted)
  & Free-form text--region grounding
  & Open-vocabulary
  & 2D only; no volumetric context \\
 
LLMSeg \cite{oh2024llm}
  & 2024
  & CT + clin.\ metadata
  & Native 3D
  & Automatic
  & Structured clinical context (stage, laterality) via LLM
  & Fixed (CTV targets)
  & Structured input only; retraining for guideline updates \\
 
Radformer \cite{rajendran2025large}
  & 2025
  & CT + clin.\ record
  & Native 3D (crop)
  & Automatic
  & Unstructured record summarized by GPT-4, embedded via PubMed-BERT; fused at each encoder stage
  & Fixed (GTV)
  & GTV only; single-institution data, no external validation \\

VoxTell \cite{rokuss2025voxtell}
  & 2025
  & Multi-mod.\ medical
  & Native 3D
  & Automatic (text-prompted)
  & Free-form clinical text drives segmentation directly
  & Open-vocabulary
  & Lacks RT-specific concepts (e.g., vol.\ margins) \\
 
\bottomrule
\end{tabular*}
\end{sidewaystable}

\begin{revision_1}
    \subsection{Search Strategy and Literature Selection} \label{sec:search}

    With the present article being designed as a narrative review, a 
    PRISMA-compliant systematic search was neither conducted nor applicable. 
    Instead, the literature selection followed a structured yet non-systematic approach to trace the major conceptual developments in segmentation for radiation oncology, from atlas-based methods to agentic architectures.

    The review was anchored in seminal works for each methodological generation (e.g., \textit{U-Net}, \textit{nnU-Net}, \textit{SAM}, \textit{CLIP}). 
    From these anchor publications, we selectively reviewed key references cited therein as well as notable subsequent work that extended or adapted them for medical or radiotherapy-specific applications. 
    Additional literature was identified through targeted searches in PubMed, Google Scholar, and arXiv using domain-specific terms (e.g., \enquote{auto-segmentation in radiotherapy}, \enquote{vision--language model segmentation}, \enquote{promptable segmentation}). No fixed date restriction was applied. However, the review emphasizes developments from approximately 2015 onward, when deep learning methods began to reshape the field, with particular attention to work published from 2023 onward.

    Publications were included if they either directly addressed segmentation in a radiotherapy context or introduced methodological advances in medical image segmentation or computer vision that substantially influenced the field. The designation of a contribution as a \enquote{milestone} was based on the authors' collective assessment of conceptual novelty, demonstrated impact on subsequent work, and relevance to radiotherapy practice. 
    We acknowledge that this selection inherently reflects the authors' expertise and perspective, which is a recognized limitation of the narrative review format.
\end{revision_1}


\section{The Previous Generation: Atlas-Based Segmentation} \label{sec:atlas}

Atlas-based methods were among the first automatic segmentation approaches in medical imaging.
These methods propagate expert-defined labels from annotated reference images to new patient data through deformable image registration \cite{cabezas2011review, rueckert2002nonrigid}.
This strategy is particularly effective for anatomical regions with limited inter-patient variability, such as the brain \cite{heckemann2006automatic}, and can delineate structures even when local image contrast is weak \cite{cabezas2011review}.

However, extensive morphological deviations, such as postoperative alterations or large tumors, can substantially impair registration accuracy \cite{iglesias2015multi}.
Multi-atlas frameworks, in which multiple atlases are registered and combined via label fusion strategies, were introduced to improve robustness against such variability \cite{iglesias2015multi}.

Atlas-based methods marked the first generation of commercially available auto-segmentation tools in radiation therapy, with Elekta \textit{ABAS} receiving FDA 510(k) clearance in 2008 \cite{han2008atlas}.
Subsequent clinical validations confirmed that using these tools substantially accelerated the contouring workflow \cite{teguh2011clinical, la2012systematic}.
These approaches were the dominant paradigm before the widespread adoption of convolutional neural networks.

\section{The Current Clinical Generation: Data-Driven but Task-Specific Segmentation} \label{sec:cnns}

\subsection{Convolutional Neural Networks}

\textit{Convolutional neural networks} (\textit{CNNs}), first conceptualized by Fukushima \cite{fukushima1980neocognitron} and later trained end-to-end by LeCun \textit{et al.} \cite{lecun1989backpropagation}, form the basis of modern computer vision and have reshaped medical image analysis for the past decade \cite{litjens2017survey}.
By applying learnable convolutional filters to local image patches, \textit{CNNs} extract hierarchical features, from low-level edges to increasingly abstract representations \cite{zeiler2014visualizing}.
While \textit{CNNs} were initially developed for image classification, segmentation requires voxel-wise predictions that preserve fine-grained spatial information, leading to the development of fully convolutional architectures that produce dense output maps directly from full-resolution images \cite{long2015fully}.

\subsection{3D U-Nets as the Radiotherapy Workhorse}

\begin{figure}[htbp]
  \centering
  \includegraphics[width=\textwidth]{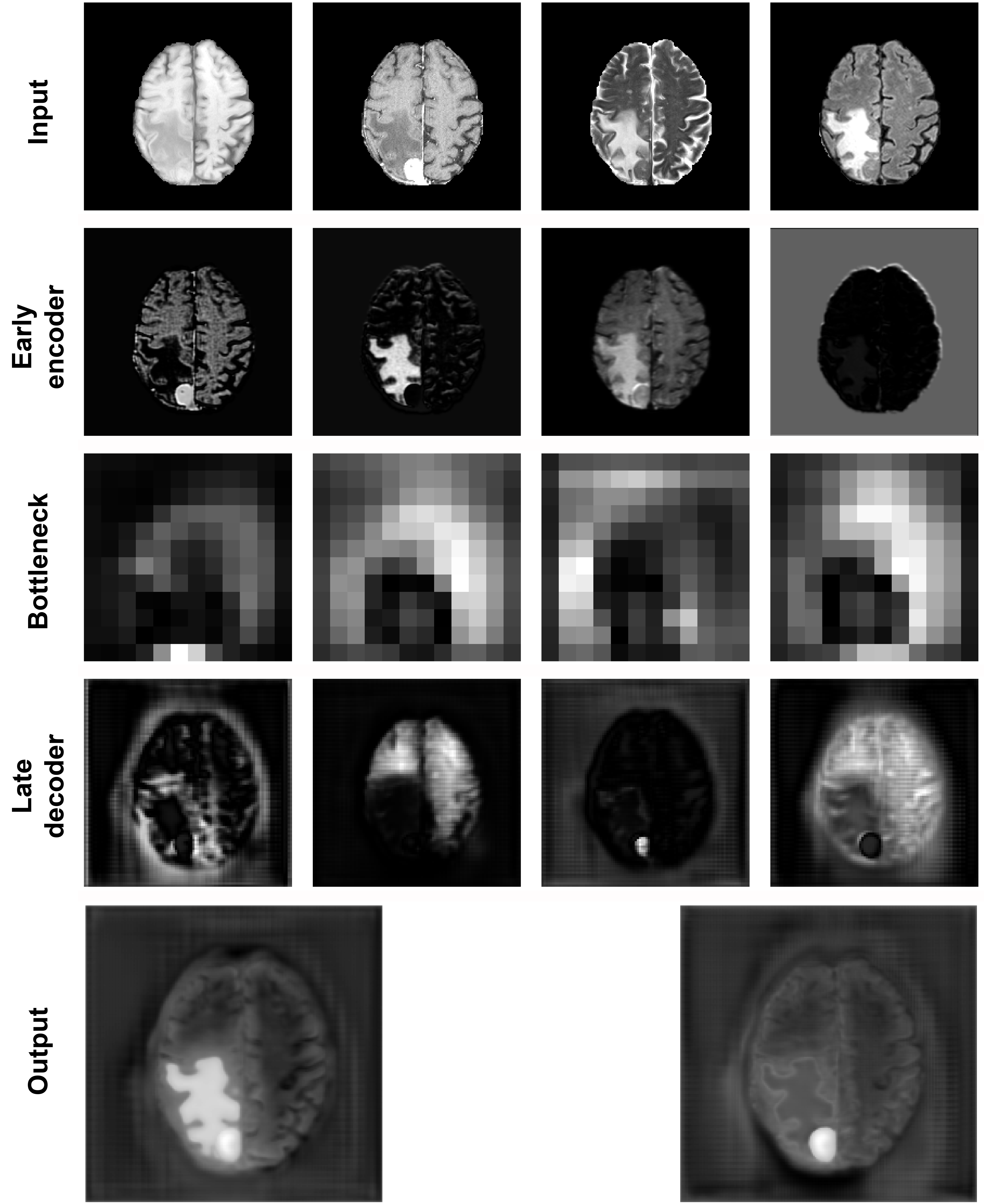}
  \caption[Visualization of 3D U-Net activations.]{\textbf{Visualization of 3D U-Net activations.} 
  Example activation maps from multiple stages of a 3D U-Net with 4 input channels and 2 output channels trained for brain metastasis and edema segmentation \cite{buchner2023development}. 
  The top row shows the four MRI input sequences (T1 native, T1 contrast enhanced, T2, and T2 Fluid Attenuated Inversion Recovery) fed into the network simultaneously. 
  Subsequent rows show representative activation maps at different network stages.
  Because all input sequences are processed jointly, individual activations do not correspond to a single sequence. 
  For each stage, four selected maps are shown for illustration.
  The total number of feature maps varies across stages. 
  In the output row, binary labels for the two output channels (whole tumor (metastasis + edema) and metastasis) are obtained by thresholding the final activation. 
  The progression from low-level edge features in early encoder layers to highly abstract representations at the bottleneck, and back to spatially precise predictions in the decoder, is clearly visible.}
  \label{fig:activations}
\end{figure}

A major breakthrough in medical image segmentation was the \textit{U-Net} architecture \cite{ronneberger2015u}, which established the encoder-decoder paradigm with skip connections as the dominant design for dense prediction tasks.
The encoder progressively downsamples the input to learn abstract representations, while the decoder recovers spatial resolution for pixel-wise predictions; skip connections fuse high-level semantic information with low-level spatial detail to preserve localization accuracy.
For illustrative purposes, selected layers of a \textit{U-Net} used for brain metastasis segmentation are shown in Figure \ref{fig:activations} \cite{buchner2023development}.

The \textit{U-Net} and its many variants, including three-dimensional extensions \cite{cciccek20163d}, remain the backbone of most contemporary radiotherapy auto-segmentation pipelines \cite{azad2024medical}.
Although architectural innovations such as attention mechanisms \cite{oktay2018attention} and residual designs \cite{he2016deep} have improved accuracy, the \textit{nnU-Net} framework demonstrated that much of the performance variability arises from suboptimal pipeline configuration rather than architectural novelty \cite{isensee2021nnu}.
By automatically adapting preprocessing, network depth, patch size, and training schedules to a dataset's characteristics, \textit{nnU-Net} achieved state-of-the-art performance across a wide range of benchmarks without manual intervention, establishing it as the de facto standard for medical segmentation studies.

\textit{TotalSegmentator} \cite{wasserthal2023totalsegmentator} builds on \textit{nnU-Net} to segment 104 anatomic structures in CT, achieving a mean Dice similarity coefficient (DSC) of 0.943.
Notably, its label set is limited to anatomical structures and does not include radiotherapy-specific target volumes such as GTVs or CTVs, which require clinical reasoning beyond anatomical identification.
Beyond technical performance, the study highlighted the clinical and scientific potential of large-scale automatic segmentation by applying the model to 4004 CT examinations and demonstrating age-related changes in organ volume and attenuation, including a positive correlation between age and aortic volume. 
Thus, \textit{TotalSegmentator} illustrates how robust large-label-set segmentation can enable quantitative imaging analyses at a scale that would be difficult to achieve manually.

The maturity of \textit{CNN}-based architectures also enabled the first commercially available deep learning auto-segmentation tools in radiation therapy, beginning with Mirada \textit{DLCExpert}, which received FDA 510(k) clearance in 2018 \cite{lustberg2018clinical, mirada2018fda}.
Comparative evaluations consistently demonstrated that deep learning-based tools outperformed atlas-based approaches in both geometric accuracy and correction time \cite{lustberg2018clinical, zabel2021clinical, gibbons2023clinical}.

\begin{revision_1}
    The models discussed in this section have been applied most widely to OAR segmentation, in which the target structures are anatomically defined, and the task can be fully specified from imaging alone. 
    GTV segmentation has also been addressed with task-specific \textit{CNN} architectures, although typically in disease-specific settings with dedicated training data.
    CTV segmentation, by contrast, has received comparatively little attention from purely vision-based architectures, as the target definition depends on patient-specific clinical context that is not or only indirectly encoded in the image.
\end{revision_1}

One limitation of \textit{U-Net}-based architectures is their limited ability to model long-range spatial dependencies \cite{luo2016understanding}.
Due to progressive downsampling operations in the encoder, the effective receptive field primarily increases at coarser spatial resolutions.
Consequently, fine-grained local details and global contextual information are only indirectly integrated, making it challenging to model small structures within a large anatomical context.
These limitations have motivated the integration of attention mechanisms and transformer-based architectures, which can model global interactions more explicitly.

\subsection{From Convolutions to Global Context}

In 2017, transformers revolutionized natural language processing with the introduction of the self-attention mechanism \cite{vaswani2017attention}, which was subsequently applied to computer vision with \textit{Vision Transformers} (\textit{ViTs}) \cite{dosovitskiy2020image}.
Rather than processing images with convolutional kernels, \textit{ViTs} divide an image into fixed-size patches and apply self-attention globally across all patches.
This enables the model to capture long-range spatial relationships, which convolutional networks only model implicitly.

In medical image segmentation, vision transformers have been incorporated into encoder-decoder architectures inspired by the \textit{U-Net} paradigm.
Hybrid \textit{CNN}-transformer models such as \textit{TransUNet} \cite{chen2021transunet} combine convolutional layers for local feature extraction with transformer modules for global context, while \textit{UNETR} \cite{hatamizadeh2022unetr} uses a pure \textit{ViT} encoder capable of handling volumetric 3D inputs directly.
Both architectures have demonstrated competitive performance on medical image segmentation benchmarks.

However, vision transformers typically require larger training datasets and substantially greater computational resources than \textit{CNN}-based models, as self-attention lacks the spatial inductive bias of convolutions and its memory requirements scale quadratically with sequence length ($\mathcal{O}(n^2)$).
Overall, vision transformers can be considered an evolutionary extension of deep learning-based segmentation that connects classical convolutional architectures with the growing category of large-scale foundation and multimodal models, but they have not replaced \textit{CNN}-based approaches in clinical practice.

\section{The Paradigm Shift: From Static Models to Promptable Interfaces} \label{sec:promptable}

\subsection{Foundation Models in Computer Vision}
In recent years, computer vision has seen the emergence of models trained on large-scale, highly diverse datasets.
These models can be fine-tuned or adapted for a wide range of downstream tasks.
Bommasani \textit{et al.} introduced the term \textit{foundation models} to describe these models. 
They emphasized the role of foundation models as general-purpose architectures that serve as a common basis for task-specific adaptation rather than as fully specialized solutions \cite{bommasani2021opportunities}.
\begin{revision_1}
In the following, we use \textit{multimodal} for any model that integrates more than one input type, such as imaging together with clinical text, and \textit{vision--language model} (VLM) for multimodal models that jointly process images and natural-language text.
\end{revision_1}

\subsection{Promptable Segmentation}
Promptable segmentation has its conceptual roots in interactive image segmentation methods developed since the 1990s \cite{mortensen1995intelligent}.
In these methods, user input, such as seed points, scribbles, or bounding boxes, guided the segmentation process. 
However, it only emerged as a clearly defined, scalable, and task-agnostic paradigm with the advent of large foundation models, most notably \textit{Segment Anything} (\textit{SAM}) in 2023 \cite{kirillov2023segment}.

Kirillov \textit{et al.} formalized promptable segmentation as a segmentation task conditioned on auxiliary input that specifies what should be segmented in an image. 
Prompts can take various forms, including spatial cues such as points, bounding boxes, or coarse masks.
The original \textit{SAM} paper also demonstrated the model’s potential compatibility with semantic prompts by showcasing a proof-of-concept text-to-mask capability, which was demonstrated in the publication but not included in the released model.
This abstraction decouples the segmentation task from fixed label definitions, enabling a single model to generalize flexibly across object categories, tasks, and domains without the need for retraining.

\textit{SAM 2} \cite{ravi2024sam} extends promptable segmentation from images to videos by treating an image as a single-frame video, thereby unifying both tasks within a single architecture.
Architecturally, \textit{SAM 2} augments the \textit{SAM} design with a streaming memory and a memory-attention module:
Features of the current frame attend to stored information from previous frames, including prior user prompts and model predictions.
This allows for temporally consistent mask propagation and interactive refinement across frames, improving the model's robustness to motion, appearance changes, and occlusions \cite{ravi2024sam}.

A key driver of scale in both \textit{SAM} \cite{kirillov2023segment} and \textit{SAM 2} \cite{ravi2024sam} is the use of data engines, i.e., model-in-the-loop annotation pipelines in which the model generates candidate masks that human annotators then correct, iteratively increasing dataset size and diversity while reducing manual labeling effort at the same time.

\begin{revision_1} 
    \textit{UniverSeg} \cite{butoi2023universeg} introduced a conceptually different form of prompting, which specifies the segmentation task through a small set of annotated examples rather than spatial cues. 
    Given a query image and a support set of image-label pairs, a convolutional encoder-decoder with so-called CrossBlocks transfers information from the examples to the query.
    This allows for the solution of previously unseen tasks without the need for retraining or fine-tuning. 
    \textit{UniverSeg} was trained on 53 public medical datasets spanning 16 imaging modalities and clearly outperformed few-shot segmentation baselines on held-out tasks.
    A gap nevertheless remained between \textit{UniverSeg} and task-specific \textit{nnU-Net} models (mean DSC of 0.72 vs.\ 0.84). 
    However, the model is restricted to binary segmentation of 2D slices at a resolution of $128 \times 128$ pixels, and its performance depends on the size and composition of the support set. 
    Nevertheless, example-based prompting is appealing for radiotherapy because a support set of previously approved contours could convey institution-specific contouring conventions or target definitions that are difficult to express through spatial prompts without requiring model retraining.
\end{revision_1}

\section{Bridging the Domain Gap: Medical Adaptations of Foundation Models} \label{sec:foundational}

General-purpose foundation models, such as \textit{SAM}, have been systematically evaluated on medical imaging data in zero-shot or out-of-domain settings \cite{he2023computer}.
Although these models perform satisfactorily on structures with well-defined, high-contrast boundaries \cite{zhang2023segment}, they struggle with low-contrast targets and ambiguous borders, which are common in medical imaging \cite{he2023computer}.

To address these limitations, Ma \textit{et al.} fine-tuned \textit{SAM} on over 1.5 million publicly available image-mask pairs across 10 imaging modalities \cite{ma2024segment}.
The resulting \textit{MedSAM} model substantially improves upon zero-shot \textit{SAM} performance and, in some cases, approaches the accuracy of task-specific \textit{U-Net} models.
However, \textit{MedSAM} is restricted to two-dimensional inputs and requires bounding-box prompts on each individual slice, discarding inter-slice context and making fully automatic volumetric segmentation infeasible.
A comparison with specialist \textit{nnU-Net} models on pelvic MRI structures showed that off-the-shelf \textit{MedSAM} trailed by approximately 8--15 percentage points in DSC even for well-defined, convex organs such as the bladder, rectum, and prostate, and largely failed (DSC $<$ 0.6) on disjoint, bilateral structures \cite{nguyen2025necessity}. Only after task-specific fine-tuning did the lightweight variant \textit{LiteMedSAM} reach parity with \textit{nnU-Net}, suggesting that generality alone does not yet translate into clinical reliability.

\subsection{Beyond 2D: Volumetric Promptable Segmentation}

The limitation to 2D inputs is addressed by two complementary approaches.
\textit{MedSAM 2} \cite{zhu2024medical} builds on the \textit{SAM 2} video architecture, treating individual slices of a 3D volume as video frames and employing a self-sorting memory bank that enables inter-slice information sharing from a single prompt.
However, because the volume is still represented as a series of 2D slices rather than as native 3D input, the model does not explicitly encode three-dimensional spatial relationships \cite{zhu2024medical, wang2025sam}, which may limit its ability to capture the complex spatial dependencies relevant to radiotherapy target delineation.

\textit{SAM-Med3D} \cite{wang2025sam} takes a different approach by redesigning the \textit{SAM} architecture to natively operate on 3D medical volumes, enabling a single volumetric point prompt to guide segmentation without requiring  slice-wise interaction.
A practical constraint is its reliance on small crops of 128\textsuperscript{3} voxels, necessitating sliding-window inference for clinical volumes, which increases computational cost and can introduce stitching artifacts at crop boundaries.

\subsection{Improving Interaction in 3D Volumes}

\textit{nnInteractive} \cite{isensee2025nninteractive} is a recent framework for flexible promptable 3D segmentation that, in line with the \textit{nnU-Net} philosophy \cite{isensee2021nnu}, prioritizes a strong convolutional backbone and systematic training over architectural novelty.
Trained on over 120 volumetric datasets spanning CT, MRI, and PET, it supports diverse spatial prompts, including points, bounding boxes, scribbles, and a novel lasso interaction, and provides explicit support for iterative refinement via additional positive and negative prompts.

The framework also addresses ambiguity in the intended target definition (e.g., segmenting an organ with or without an associated lesion) as a fundamental challenge of spatial prompting.
\textit{nnInteractive} handles such ambiguities by explicitly incorporating plausible ambiguous label combinations during training.
This enables the model to infer user intent through subsequent interactions rather than being restricted to a single interpretation.

From a radiotherapy perspective, spatially prompted segmentation is particularly well-suited for GTV delineation, where the target is visible on imaging but varies substantially across patients.
Because the model is not restricted to a fixed label set, it can, in principle, segment any structure identified through spatial prompts, including rare or atypical presentations that task-specific models may not have encountered during training.
The iterative refinement capability is especially relevant, as tumor boundaries are often ambiguous and may require repeated corrections.
\begin{revision_1}
    Although frameworks such as \textit{nnInteractive} are trained on datasets that include OAR structures, their primary added value over dedicated task-specific models lies in the interactive refinement of individual cases rather than in routine batch segmentation.
\end{revision_1}
However, as discussed in the following section, spatial prompting alone cannot convey the higher-level clinical concepts required for structures such as CTVs, where the segmentation target is defined by guidelines and clinical reasoning rather than just by visible anatomy.

A recent multi-institutional study \cite{julius2026gtv} evaluated \textit{SAM 2}, \textit{MedSAM 2}, and \textit{nnInteractive} for GTV segmentation on MRI-Linac data across five tumor sites.
Averaged over all prompt configurations, \textit{nnInteractive} achieved the highest median DSC (0.75), followed by \textit{MedSAM 2} (0.70) and the untuned \textit{SAM 2} (0.54).

\section{From Anatomy to Semantic Concept: Vision--Language Models in Radiotherapy} \label{sec:vlms}

\subsection{Conveying Intent via Language}

Although \textit{nnInteractive} and other spatially prompted segmentation models show that geometric user inputs can effectively guide segmentation, these systems only use spatial cues, such as points, boxes, or scribbles, for prompting.
In clinical practice, however, segmentation tasks rarely involve delineating a single instance.
Instead, they usually require segmenting all instances of a clinically meaningful category, such as all distinct tumor lesions, or affected lymph nodes.

\begin{figure}[htbp]
  \centering
  \includegraphics[width=\textwidth]{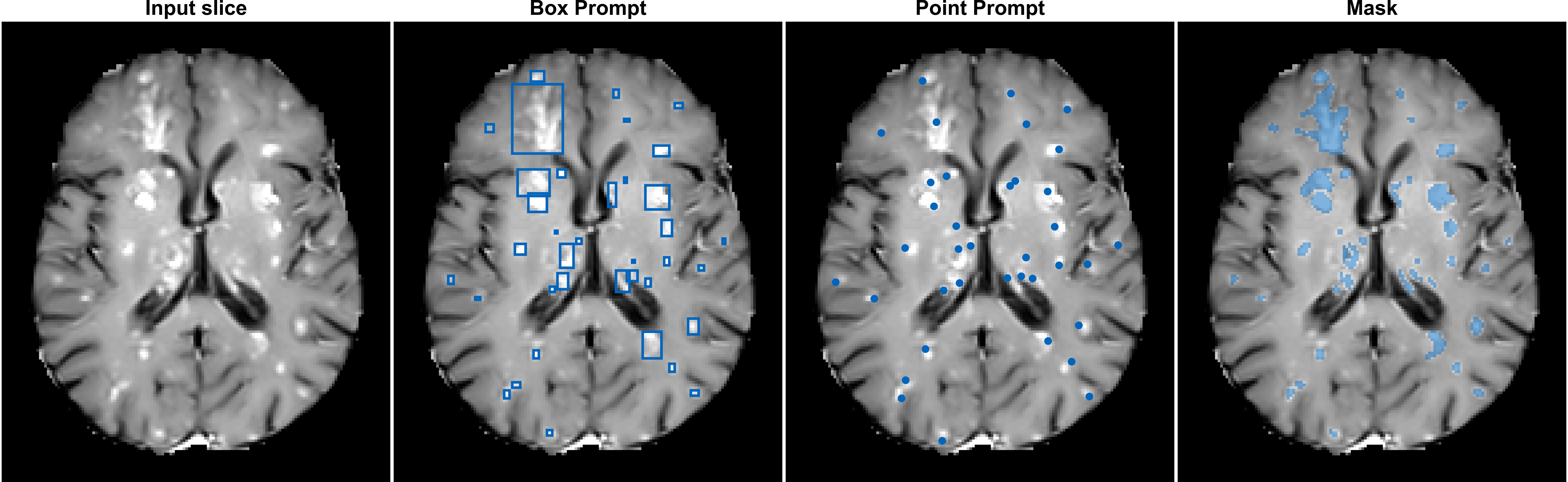}
  \caption[Spatial prompting in brain metastasis segmentation.]{\textbf{Spatial prompting in brain metastasis segmentation \cite{moawad2024brain}.} 
  Especially in segmentation tasks with multiple instances of the same target, such as brain metastasis segmentation, using individual spatial prompts like point prompts or bounding boxes for each lesion is inefficient, as patients may present with dozens of metastases that would each require separate annotation.}
  \label{fig:prompts}
\end{figure}

In such scenarios, prompting each instance individually becomes inefficient and difficult to scale, particularly when the targets are numerous (Figure~\ref{fig:prompts}), when they are densely distributed, or when they are defined as an anatomical concept rather than a discrete object, such as a lymphatic drainage region.
Language provides a more flexible interface for task specification and enables users to describe target structures semantically and to convey aspects of clinical intent.
Here, the target is specified through textual inputs rather than by repeatedly placing spatial prompts for each instance.

\subsection{Vision--Language Alignment}

\subsubsection{CLIP-based Approaches}

For language to guide segmentation, image features and text need to be brought into correspondence.
One influential approach is to map both modalities into a shared representation space.
An important milestone in this direction was \textit{CLIP} \cite{radford2021learning}, which demonstrated that image and text representations can be aligned at scale through contrastive pretraining on 400 million image-text pairs. 
Instead of learning a fixed classifier over a closed label set, \textit{CLIP} maps images and natural-language descriptions into a shared embedding space. 
Because image and text embeddings occupy the same space, the model can be queried without retraining on a fixed label set.

An early step towards applying this concept to medical image segmentation was \textit{CLIP-driven Universal} \cite{liu2023clip}. 
This model incorporates \textit{CLIP}-derived text embeddings of class names into a universal abdominal CT segmentation framework. 
In this way, labels are no longer represented as unrelated one-hot identifiers, but as semantically structured embeddings. 
This allows the model to exploit relationships between anatomically related structures and to learn a more structured feature space. 
The authors achieved top-ranked results on the Medical Segmentation Decathlon leaderboard with this design \cite{antonelli2022medical}.

While \textit{CLIP-Driven Universal} demonstrates the value of language-grounded label representations for structured 3D segmentation, it does not support free-form text prompts during inference. 
Instead, it uses \textit{CLIP}-derived embeddings of a predefined set of class labels, so language serves as a semantic prior for fixed categories rather than as an interface for arbitrary user queries.

\textit{SegVol} \cite{du2024segvol} moves language from a training-time prior to an inference-time input.
Built on a \textit{ViT} encoder \cite{dosovitskiy2020image}, the model accepts spatial prompts such as points and bounding boxes alongside semantic text prompts encoded via a frozen \textit{CLIP} text encoder.
The authors demonstrate that jointly providing a bounding box and a textual description yields the highest segmentation accuracy, and that adding a text prompt improves the average DSC by 3.33 percentage points over bounding boxes alone.
This can resolve the spatial ambiguity inherent to geometric prompts when multiple structures overlap.

To handle high-resolution volumetric data efficiently, \textit{SegVol} introduces a zoom-out–zoom-in inference strategy. 
Rather than densely covering the entire volume via sliding-window inference, the model first performs a global pass at reduced resolution to identify the approximate region of interest.
This is followed by a targeted high-resolution pass over the localized region to reduce computational cost.
Although trained on CT data, \textit{SegVol} demonstrates partial generalization to MRI. 

However, because the \textit{CLIP} encoder remains frozen and was pretrained on natural image-text pairs, its capacity to represent specialized clinical or anatomical terminology may be limited.
Domain-adapted alternatives such as \textit{BiomedCLIP}, which was pretrained on 15 million biomedical image-text pairs, have demonstrated substantially improved performance on biomedical vision--language tasks compared to the general-domain \textit{CLIP} \cite{zhang2025multimodal}.

\subsubsection{Free Text Prompting}

\textit{BiomedParse} \cite{zhao2025foundation} takes a further step toward genuine text-prompted segmentation. 
It is a biomedical vision--language foundation model that can perform joint segmentation, detection, and recognition across multiple imaging modalities.
It also has the ability to segment objects in images directly from free-form text prompts. 
By associating textual entity descriptions with image regions, \textit{BiomedParse} demonstrates the feasibility of specifying segmentation targets semantically rather than through explicit spatial prompts.

Across several benchmarks, \textit{BiomedParse} outperformed prior promptable baselines such as \textit{SAM} and \textit{MedSAM}, with particularly strong gains for irregularly shaped objects. 
These results highlight the potential of language as a flexible interface for defining segmentation targets beyond purely spatial prompts. 
However, the original model is limited to two-dimensional inputs.
For volumetric modalities such as CT and MRI, volumes are processed as stacks of in-plane slices rather than as native 3D inputs. 
This restricts the model's ability to exploit the full volumetric context, which is particularly important in radiotherapy planning. 
Furthermore, the demonstrated prompts remain closer to entity-level region grounding than to richer radiotherapy-specific clinical instructions.
Consequently, the model does not directly support the explicit modeling of higher-level clinical intent, treatment guidelines, or planning-specific concepts.

The restriction to 2D data was subsequently addressed by the successor 
model \textit{BiomedParse-V} \cite{zhao2025biomedparse}. 
Using fractal volumetric encoding, the authors hierarchically compress volumetric data into a compact 2.5D representation that preserves inter-slice context while remaining compatible with 2D RGB foundation model weights. 
In addition, an independent segmentation discriminator filters slices in which the prompted structure is absent, thereby reducing false-positive predictions. 
With these modifications, \textit{BiomedParse-V} achieved state-of-the-art performance across several 3D biomedical segmentation benchmarks, surpassing supervised 3D models.
However, like its predecessor, \textit{BiomedParse-V} does not incorporate radiotherapy-specific concepts such as target volume margins or guideline-driven contouring, so the advance remains architectural rather than domain-specific.

\subsection{Improving Alignment by Repeated Cross-modal Fusion}

Unlike earlier vision--language segmentation approaches that typically fuse language information at a single, late decoding stage, \textit{VoxTell} employs a multi-stage vision--language fusion strategy across multiple levels of the decoder hierarchy \cite{rokuss2025voxtell}.
This enables progressive alignment of the language prompt and spatial information throughout the network, rather than relying on a single semantic fusion step.

Architecturally, \textit{VoxTell} is based on a convolutional \textit{U-Net} backbone for both the encoder and decoder.
Transformer components are deliberately restricted to language and prompt processing, where they encode free-form textual descriptions and mediate cross-modal interactions.

\begin{figure}[htbp]
  \centering
  \includegraphics[width=\textwidth]{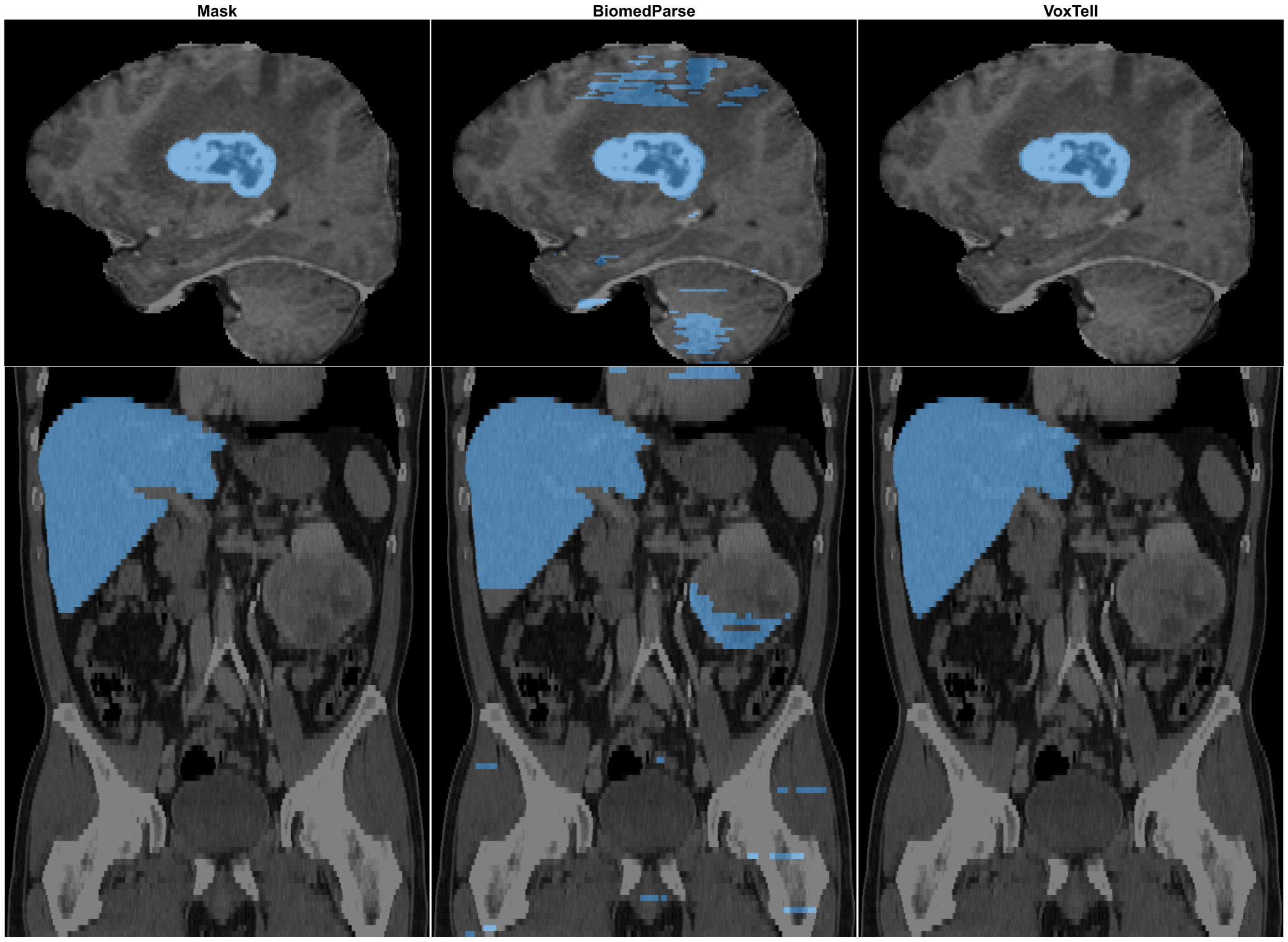}
  \caption[Comparison of 2D and 3D text-prompted segmentation.]{\textbf{Comparison of 2D and 3D text-prompted segmentation.} 
  Top row: sagittal view of a brain metastasis (GTV) segmented with the text prompt \enquote{tumor core} on MRI. 
  Bottom row: coronal view of the liver (OAR) segmented with the text prompt \enquote{liver} on CT.
  In both cases, \textit{BiomedParse} (center) segments each axial slice independently, and the out-of-plane reconstruction reveals staircase-like artifacts and scattered false positives. 
  \textit{VoxTell} (right), which processes the volume natively in 3D, produces spatially coherent segmentations that closely match the expert reference masks (left). 
  Data from an in-house brain metastasis dataset \cite{buchner2023development, buchner2023identifying} and the TotalSegmentator dataset \cite{wasserthal2023totalsegmentator}.}
  \label{fig:2D_3D}
\end{figure}

A major practical advantage of \textit{VoxTell} is that it operates directly on 3D data.
This addresses a limitation of many earlier promptable segmentation models like the original \textit{BiomedParse} that are restricted to two-dimensional inputs.
Processing volumes slice-by-slice can produce spatial inconsistencies, including staircase-like artifacts and false positives, as shown in the comparison in Figure \ref{fig:2D_3D}.
By conditioning the decoding process on language at multiple spatial scales, \textit{VoxTell} supports flexible, text-driven target definitions while maintaining spatial coherence across slices.
The authors show that \textit{VoxTell} can handle clinical language similar to what is found in radiology reports as input.

Notably, the authors included a radiotherapy-specific evaluation in their study, testing \textit{VoxTell}'s ability to generate GTV segmentations for stereotactic lung radiotherapy using clinical language prompts extracted directly from radiology reports. 
With complex prompts such as \enquote{spiculated carcinoma in the left upper lobe with pleural contact}, the model achieved DSC of approximately 0.5, substantially outperforming VLMs such as \textit{SegVol} on the same task.

In our own testing with an in-house brain metastasis dataset \cite{buchner2023development, buchner2023identifying}, however, we found that while \textit{VoxTell} is capable of generating precise segmentations of the metastases and achieved DSCs comparable to expert models, it does not understand essential concepts in radiotherapy like adding volumetric margins.
Therefore, even though the model demonstrates broad anatomical understanding, this cannot be equated with meaningful understanding of clinical concepts in radiation oncology.

Across the models discussed in this section, language has progressed from a training-time regularizer over fixed label sets, as in \textit{CLIP-driven Universal}, to an inference-time interface that maps free-form clinical text directly to volumetric segmentation masks, as in \textit{VoxTell}.
This distinction is important for radiotherapy, where segmentation targets are often defined by complex, domain-specific instructions that a model must interpret at inference time rather than encode implicitly during training.

\subsection{Automated CTV definition}

Unlike the anatomical structures targeted by the models discussed above, CTVs are often not visually identifiable in imaging, which makes multimodal integration essential. 
\begin{revision_1}
    While the VLMs discussed in the preceding section represent a qualitative advance for GTV segmentation by enabling semantic target specification through free-form text, their utility for CTV delineation remains limited: as demonstrated by our evaluation of \textit{VoxTell}, the previously discussed VLMs lack an understanding of radiotherapy-specific concepts such as volumetric margin expansion and do not incorporate clinical guidelines or patient-specific metadata into the segmentation process.
\end{revision_1}

\textit{LLMSeg} \cite{oh2024llm} aims to automate CTV segmentation, primarily validated for breast carcinoma with an extension to prostate carcinoma.
The authors use structured clinical input, such as tumor stage or laterality, encoded via a large language model (LLM) as an additional input. 
Adding the textual information improved performance compared to vision-only models, especially in the external validation set and in data-scarce settings. 

The authors demonstrated that modifications to the text input, e.g., upstaging the nodal status, change the corresponding output, e.g., by including regional lymph drainage, suggesting that the model encodes clinically meaningful relationships rather than superficial correlations. 

\begin{revision_1}
    A conceptually related approach is \textit{Radformer} \cite{rajendran2025large}, which was developed for head and neck cancer using the public \textit{RADCURE} dataset \cite{welch2024radcure}. 
    Unlike \textit{LLMSeg}, clinical information is not supplied as predefined structured fields. 
    Instead, GPT-4 is prompted to extract tumor-related information from the clinical record, which is then embedded with PubMed-BERT.
    These language features are fused with the image features at multiple stages of a hierarchical Swin Transformer encoder, comparable to the multi-level fusion strategy of \textit{VoxTell}. 
    In a held-out test set of 597 patients, adding language features improved the mean DSC from 0.70 to 0.76 compared to the same architecture without language fusion. 
    However, \textit{Radformer} was trained and evaluated exclusively on GTVs from a single institution without external validation. 
    Therefore, its results indicate that clinical context can improve segmentation even of visible targets, but do not yet address the guideline-driven reasoning required for CTV delineation.
\end{revision_1}

A limitation of \textit{LLMSeg} is its restriction to structured clinical data, therefore not allowing free-form natural-language interaction\begin{revision_1}.
Although \textit{Radformer} accepts unstructured clinical records, these are condensed into a fixed summary prior to inference rather than serving as an interactive interface\end{revision_1}. 
Furthermore, \begin{revision_1}in both approaches,\end{revision_1} clinical guidelines are not incorporated explicitly during inference, but instead learned implicitly during training.
As a consequence, updates to contouring guidelines would likely require retraining of the model.

\subsection{Improving Segmentation Quality by Including Uncertainty} \label{sec:uncertainty}

The quantification of segmentation uncertainty is particularly relevant for CTV delineation, where the target is often only indirectly inferable from imaging.
In clinical practice, substantial interobserver variability persists even among experienced radiation oncologists. 
Models that output only a single deterministic mask can obscure this ambiguity and bias raters during contour review. 

Uncertainty estimation methods have already been integrated into deep learning-based segmentation frameworks. 
For example, Balagopal \textit{et al.} used Monte Carlo dropout \cite{gal2016dropout} to generate 95\% confidence bands for postoperative prostate CTV segmentation, thereby making spatial variation in model confidence explicit \cite{balagopal2021deep}. 
In their framework, these uncertainty bounds were intended to highlight regions that may require closer physician review, especially in a task where no directly visible ground-truth target exists. 
This can lead to time-savings during manual corrections by influencing the decision-making and quality perception of radiation oncologists \cite{rogowski2025impact}.

Integrating similar uncertainty-aware mechanisms into language-guided CTV models would be a logical next step, for instance, by defining sources of Bayesian variation based on the supply or processing of language, or its abstractions and features, within such models. 
Such methods and respective standardizations of their application can help radiation oncologists focus their review on anatomically or semantically ambiguous regions and may improve the consistency and transparency of the final approved contour.


\section{Agentic Artificial Intelligence in Radiation Oncology} \label{sec:agentic}

\subsection{From Static Tools to Autonomous Agents}

All of the systems discussed so far are passive tools that depend on a human user to select them, prompt them, interpret their output, and decide on the next step. 
In a recent Comment in \textit{The Lancet}, Zou and Topol argue that medicine is at the beginning of a conceptual shift from treating AI as passive tools toward deploying it as agents \cite{zou2025rise}.
This enables the models to take initiative, for instance, within a defined clinical task rather than wait for instruction.

Architecturally, agentic systems are typically built around an LLM that maintains an overview of the available data, imaging studies, and downstream tools, and that can invoke specialist modules as needed, while acting as a flexible controller that reasons about which tools to invoke and in what order \cite{li2026agentic}. 
Specialist modules in this setting could include the segmentation models discussed in the previous sections, registration tools, report parsers, monitoring routines \cite{rafael2025agent}, or guideline retrieval components. 

Such an approach has been demonstrated by Erdur \textit{et al.} for the analysis of neuro-radiological imaging \cite{erdur2026agentic}. 
Their system orchestrates off-the-shelf tools for preprocessing, tumor segmentation, and volumetric analysis to autonomously execute multi-step workflows, including longitudinal response assessment, without any training or fine-tuning of the underlying LLM. 
This illustrates how LLMs can coordinate complex volumetric imaging tasks despite lacking the ability to process 3D imaging data directly.

\subsection{Proof of Concept in Radiotherapy Planning} 

A proof of concept that this paradigm is feasible in radiation oncology, albeit for treatment planning rather than segmentation, is \textit{GPT-RadPlan} \cite{liu2025automated}. 
The system is built around three modules that are all powered by GPT-4V \cite{openai2023gpt4v}: an evaluation module that inspects the current dose distribution and the corresponding DVH table through dedicated image and DVH sub-experts, a memory module that retains previous plans together with their optimization parameters, and a planner module that translates the evaluator's feedback into concrete adjustments of objective weights and dose constraints. 
\textit{GPT-RadPlan} does not generate plans from scratch but instead wraps an existing inverse planning system: an outer loop driven by the planner module iteratively reshapes the optimization parameters of the plans produced by the conventional inner optimizer.
Domain knowledge is supplied through in-context learning, using three approved reference plans together with their optimization settings, without any fine-tuning of GPT-4V. 
In a preliminary evaluation on 17 prostate and 13 head and neck cases, \textit{GPT-RadPlan} produced plans that were competitive with or, in several metrics, exceeded the corresponding clinical plans created by human experts.

\subsection{Towards Agentic CTV Segmentation} \label{sec:agents} 

A similar agentic workflow can be imagined for the automatic creation of CTVs.
As discussed previously, a key limitation of \textit{LLMSeg} is that it learns contouring guidelines implicitly during training. 
Therefore, periodic updates to the guidelines necessitate retraining the model.

In an agentic system, this could be addressed by treating the guidelines as an external knowledge base.
Utilizing retrieval-augmented generation (RAG) \cite{lewis2020retrieval}, suitable documents are selected from such a knowledge base and supplied to a language model during inference.
This allows models to base their outputs on supplied documents, which are integrated into the context of the underlying agent model.

\begin{figure}[htbp]
  \centering
  \includegraphics[width=\textwidth]{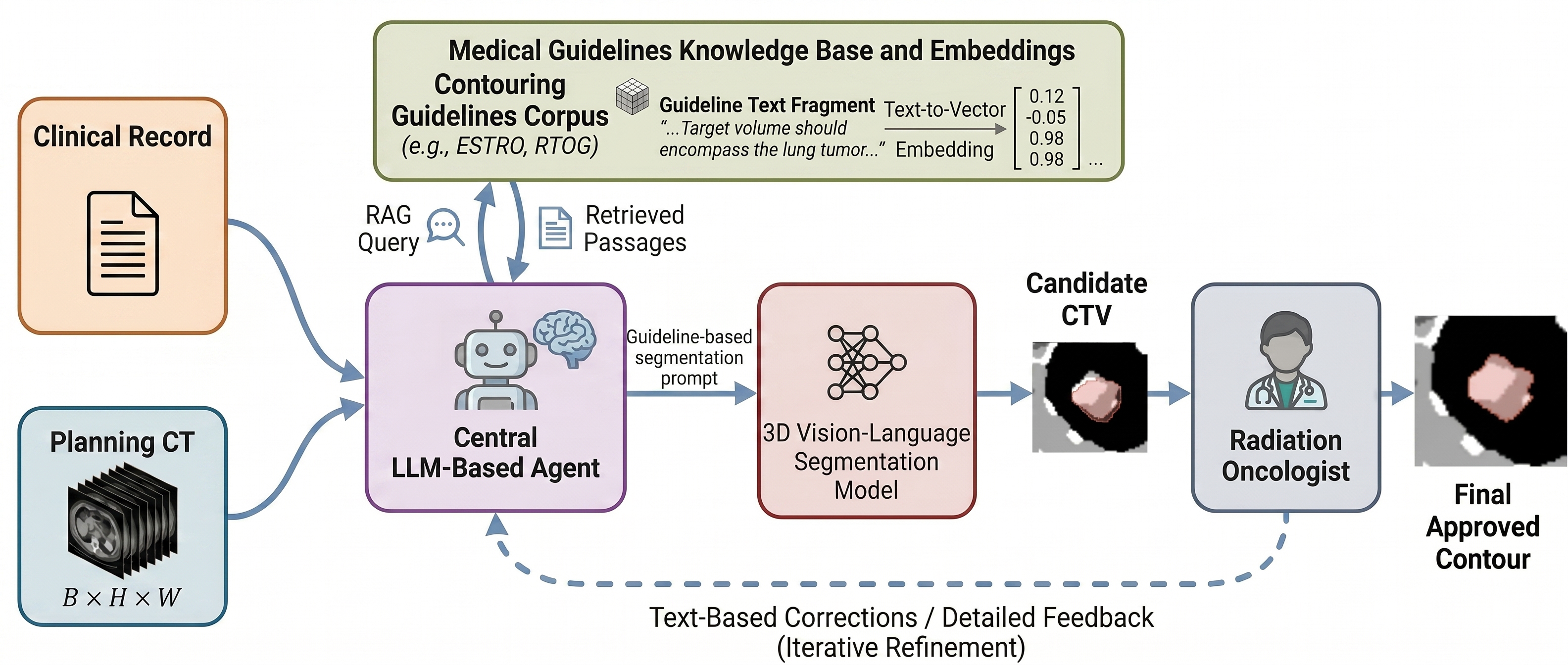}
  \caption[Possible future workflow of agentic segmentation system.]{\textbf{Possible future workflow of agentic segmentation system.} 
  The central LLM-based agent ingests the planning CT and clinical record, retrieves relevant guideline passages via RAG, and generates segmentation prompts for a 3D VLM. 
  The radiation oncologist approves the candidate CTV and can provide text-based corrections via an iterative refinement loop.}
  \label{fig:agentic_workflow}
\end{figure}

Concretely, an agentic CTV system could ingest a planning CT together with the clinical record, retrieve relevant clinical guidelines for the current indication, and formulate segmentation instructions derived from the selected parts of the guidelines.
These instructions can be used as prompts for VLMs, which are then translated into 3D segmentations of CTVs.
Figure \ref{fig:agentic_workflow} illustrates a possible workflow of such an agent-based system.

In this setup, the human role shifts from operating individual segmentation models toward supervising and verifying an automated contouring workflow. 
Because language is utilized as the primary interface throughout the pipeline, integrating an interactive feedback loop is essential.
This way, the segmentations could be interactively modified by human experts using additional textual inputs in clinical language or spatial prompts. 

\subsection{Challenges and Implications}

\subsubsection{Current Technical Limitations}

\begin{revision_1}
Realizing such a framework requires careful attention to failure modes that are specific to agentic, multi-step systems. 

First, retrieval can return outdated, incomplete, or non-applicable guideline passages, silently grounding a contour in the wrong standard or a standard not relevant to a local community.

Second, the language model may justify a contour with a rationale that is not supported by the retrieved sources. Even with retrieval-augmented
generation, LLMs remain prone to hallucinations in medical question answering~\cite{vladika2025improving}, and a fluent but unsupported justification may lend a contour unwarranted credibility.

Third, the segmentation model may not act on all parts of its instruction: In prompt-perturbation experiments with the zero-shot VLM \textit{VoxTell}, non-small cell lung cancer (NSCLC) GTV segmentations depended chiefly on the anatomical location stated in the prompt and hardly on histology or tumor stage~\cite{pai2026prompt}. While appropriate for a primary GTV, this suggests that clinical attributes that should alter a target, such as
nodal stage for elective nodal volumes, may be silently disregarded.

Fourth, because modules are chained, an early error can propagate and compound downstream or be masked by a compensating error, so that a plausible final contour reveals little about the correctness of intermediate steps, as revisited in Section~\ref{sec:agent_eval}. 

These risks reinforce automation bias (see Section~\ref{sec:risks}), as the reviewing physician is presented not only with a complete contour but also with an apparently reasoned, and potentially convincing, justification. 
Finally, attributing an incorrect contour to prompting, retrieval, reasoning, or segmentation is complex, which complicates quality assurance and can leave an assignment of responsibility unclear. 
Logging of intermediate steps for audit, exposure of model uncertainty (see Section~\ref{sec:uncertainty}), the traceability mechanisms discussed below, and mandatory physician verification should therefore be considered preconditions for clinical use rather than optional safeguards. At the same time, a rigorous review would require substantial time, diminishing potential gains in efficiency. 
\end{revision_1}

Beyond these system-level failure modes, several technical gaps remain before agentic segmentation can be adopted in clinical practice.
At the moment, segmentation VLMs do not have a reliable understanding of radiotherapy-specific concepts, such as volumetric margin expansion, as is shown by our own evaluation of \textit{VoxTell} and confirmed by systematic benchmarking of vision--language and \textit{SAM}-based models across CT and MRI tasks \cite{erdur2025benchmarking}.  

Moreover, no model has yet been shown to delineate guideline-defined target subvolumes, such as specific lymphatic drainage levels, from textual guideline descriptions.

RAG pipelines have been explored for clinical guidelines in radiology \cite{komenda2026development}, but their application to the structured, highly specific language found in radiotherapy contouring guidelines remains unexplored.
It remains unclear how reliably such systems can distinguish between applicable and non-applicable guideline passages for a given clinical scenario, and whether detailed, often unstructured descriptions of target volumes with varying nomenclatures can be safely transformed into adequate 3D segmentation.

Mechanisms to trace individual components of an LLM's output to their source statements in the knowledge base, such as atomic fact checking \cite{vladika2025improving}, may need to be employed to enable clinicians to verify the guideline basis of contouring decisions.

\subsubsection{Potential Benefits}

If these limitations can be addressed, the potential benefits of agentic therapy planning workflows are substantial.
As the radiotherapy workforce is growing more slowly than the global incidence of cancer, and therefore the demand for radiotherapy treatment \cite{zhu2024global, petit2025increasing}, it can be assumed that there will be a shortage of treatment capacity in the future.
Efficiency gains from automated contouring can help to maintain and increase this capacity.

Increased contouring speed could also support emerging treatment paradigms such as online adaptive radiotherapy, in which the treatment plan is recalculated before each fraction to account for daily changes in patient anatomy and tumor characteristics \cite{kalinauskaite2025online}. 
This workflow requires new OAR and, in some cases, CTV segmentations to be generated. 
Automated segmentation systems could help produce these contours within the narrow time constraints of an online adaptive workflow.

Furthermore, directly translating contouring guidelines into target volume definitions could promote standardization across institutions \cite{bollen2023impact}.
If this reduces the interobserver variability that currently affects the delineation of CTVs, improvements in treatment quality may follow, potentially contributing to better oncological outcomes and reduced treatment-related toxicity.
This development can particularly enhance workflows and outcomes in small centers lacking experience with certain tumor entities \cite{peters2010critical}, although these presumed effects must be demonstrated in prospective studies.

For individual radiation oncologists, shifting the majority of the segmentation burden to automated systems can free up time and attention for tasks such as patient interaction and advanced treatment planning. 

The modular architecture of agentic systems means that their capabilities can be extended beyond contouring by integrating additional specialist tools.
For instance, incorporating deformable image registration enables reirradiation analysis with deformable EQD2 maps. 
Coupling the system with tumor detection systems supports the automated identification of targets, such as progressive brain metastases.
Similarly, integrating radiomic feature extraction and longitudinal imaging analysis with contouring modules within the same framework enables lesion characterization and treatment response assessment.
Once the core agentic infrastructure is in place, \begin{revision_1}new capabilities could be added by integrating further modules, each of which would require validation.\end{revision_1}

\subsubsection{Risks} \label{sec:risks}
These potential benefits, however, must be weighed against equally important risks.
A progressive shift toward fully automated 
planning pipelines may lead to deskilling of radiation oncology personnel if manual contouring expertise is no longer routinely practiced \cite{callens2024full}.
This is especially concerning for complex cases where current automated systems are most likely to fail and human expertise is most critical.

Furthermore, AI models learn patterns from their training data. 
If there are structural deviations from the guidelines or institution-specific practices in that data, these biases can be propagated and amplified when the model is deployed across institutions, directly influencing the target volume segmentation.

Finally, the widespread integration of automatic contouring systems creates new pitfalls in human-AI interaction. 
The tendency of clinicians to accept AI-generated outputs with insufficient scrutiny is a well-documented phenomenon known as automation bias \cite{lyell2017automation, mackay2025royal, nealon2022using}. 
This is particularly relevant in this context, where the contours presented to the reviewing physician already appear complete and plausible.
The uncertainty-aware mechanisms discussed in Section \ref{sec:uncertainty} could serve as a partial countermeasure by directing the reviewer's attention to regions where the model's confidence is low, thereby discouraging uncritical acceptance of the complete output.

\section{Towards New Paradigms for Next-Generation Evaluation} \label{sec:new_paradigm}
\begin{revision_1}
\subsection{Performance versus Flexibility: Dimensions of Paradigm Comparisons}
Comparing models across the paradigms described in Sections~\ref{sec:atlas}--\ref{sec:agentic} is an increasingly multi-dimensional problem. 
Beyond segmentation accuracy, a model may be assessed on uncertainty estimation, prompt adherence, linguistic and spatial flexibility, generalizability to unseen tasks and cases, computational efficiency, and, in line with consensus guidelines for trustworthy medical AI, robustness, interpretability, and fairness across patient populations~\cite{lekadir2025future}. 
If these dimensions are conflated, apparent methodological progress may not translate into clinical benefit. 
We therefore examine four claims commonly made for VLMs: that they improve on task-specific baselines in (a)~segmentation quality, (b)~interface flexibility, (c)~generalizability across tasks, anatomical sites, and imaging modalities, and (d)~uncertainty quantification. 
The literature and the architectural analysis in Sections~\ref{sec:promptable}--\ref{sec:agentic} support (b) and, in part, (c)~\cite{zhao2025sat,rokuss2025voxtell}, whereas (d) remains largely untested for language-guided models (Section~\ref{sec:uncertainty}).

For (a), the evidence remains equivocal. Across 72 public datasets, a text-prompted universal model matched, but did not exceed, per-dataset \textit{nnU-Net} models~\cite{zhao2025sat}. 
In an independent cohort of 93 NSCLC radiotherapy cases, zero-shot \textit{VoxTell} was statistically indistinguishable from, but numerically below, a fine-tuned \textit{nnU-Net} (mean DSC 0.61 vs.\ 0.69), while other zero-shot VLMs performed markedly worse (DSC $\leq$ 0.25)~\cite{pai2026prompt}. 
For glioma subregions, text-refined \textit{VoxTell} performed similarly to a T1c-only \textit{nnU-Net} on the internal test set (mean DSC 0.80 vs.\ 0.79)~\cite{eidex2026textrefine}. 
On pooled cross-dataset evaluation, it exceeded \textit{nnU-Net} (0.55 vs.\ 0.49), but pretrained \textit{VoxTell} without correction instructions already reached 0.53, and \textit{SAT-Pro}~\cite{zhao2025sat} showed a similar advantage. 
The transfer advantage was therefore not specific to instruction-guided refinement. 
Correct instructions, derived from the reference contours, added only a small further gain (0.02--0.05 DSC) over the pretrained, blank-, and contradictory-instruction conditions. 
Whether the remaining advantage over \textit{nnU-Net} reflects large-scale pretraining, text conditioning, or prior exposure to the public test cohorts cannot be resolved from the study design. 
Consistently, task-specific \textit{nnU-Net} outperformed off-the-shelf \textit{MedSAM}~\cite{ma2024segment} on pelvic MRI OARs by 0.15 mean DSC for the bladder and 0.08 for the rectum, and by 0.37--0.64 for disjoint, bilateral structures~\cite{nguyen2025necessity}.

Whether these gaps narrow, close, or reverse as VLMs are trained on larger, radiotherapy-specific datasets remains open. 
For now, the case for these models does not rest on surpassing \textit{nnU-Net} on image-visible structures, e.g. in selected cases where clinical information may inform segmentation quality. 
Rather, it rests on segmenting arbitrary or unseen structures without retraining and, most importantly for radiotherapy, on addressing guideline-defined targets such as the CTV, which image-only models cannot express.

Settling (a) is further hampered by the current evidence base. 
Same-test-set comparisons remain scarce, and evaluations are commonly run by the model developers themselves. 
Multi-institutional evaluations often rank promptable models against one another without including a fully automatic, task-specific baseline~\cite{julius2026gtv}. 
Moreover, most studies are limited to anatomical structures or GTVs, report geometric metrics only, and may be confounded by overlap between public pretraining corpora and test data. 
Resolving (a) therefore requires a standardized radiotherapy benchmark built on data withheld from all public training corpora. 
Such a benchmark should provide fixed splits, harmonized reference standards, separate tracks for automatic and prompted segmentation under a fixed prompt protocol, and a common set of geometric, effort-aware, dosimetric, and clinically validated endpoints.
\end{revision_1}

\subsection{The Limitations of Geometric Evaluation}

The majority of the aforementioned auto-segmentation studies report performance using the DSC, which quantifies volumetric overlap between a predicted label and a reference segmentation mask. 
While widely understood, DSC has well-recognized limitations as a clinical surrogate \cite{maier2024metrics}. 
It is insensitive to local surface deviations: a model can achieve high volumetric agreement while introducing boundary errors at clinically critical interfaces, such as the junction between a target volume and an adjacent OAR, requiring manual intervention and correction.

Surface-based metrics address this shortcoming partially by evaluating agreement at the contour boundary rather than across the entire volume \cite{Huttenlocher1993ComparingIU, maier2024metrics}. 
The 95th-percentile Hausdorff distance (HD95) captures the worst-case surface deviation after excluding outliers, while the mean surface distance (MSD) summarizes average boundary disagreement.
These metrics are more sensitive to the localized errors, which typically require manual intervention and have therefore become standard additions to DSC in recent segmentation challenges and benchmarking studies \cite{maier2024metrics}.
Nevertheless, surface-based metrics treat all boundary deviations symmetrically, regardless of their anatomical location or clinical consequence. 
For example, an error at a target-OAR boundary receives the same penalty as an equivalent deviation in a low-risk region.

\begin{revision_1}
Moreover, effort-aware metrics can serve as proxies for contour-editing workload.
The surface DSC measures the proportion of the combined predicted and reference surface that lies within a specified tolerance of the opposite surface, so that organ-specific acceptable deviations are disregarded~\cite{nikolov2021clinically}. 
Termed normalized surface distance (NSD) in Metrics Reloaded, it is preferred when a tolerance for imprecise reference annotations is desired~\cite{maier2024metrics}. 
However, its aggregate score does not reflect the clinical importance of the location of an error. The added path length (APL) measures the length of contour that must be added or redrawn to obtain a clinically accepted contour~\cite{vaassen2020evaluation}. Vaassen \textit{et al.}~\cite{vaassen2020evaluation} studied thoracic OARs in 20 lung cancer patients, where APL correlated most strongly with absolute contour-adaptation time ($R=0.87$), followed by surface DSC ($R=-0.69$) and the Hausdorff distance ($R=0.64$), whereas volumetric DSC correlated weakly ($R=-0.32$). For relative time saving, however, APL correlated only weakly ($R=-0.38$)~\cite{vaassen2020evaluation}. 
These findings support the use of effort-aware metrics as proxies for editing workload, whereas estimating net clinical time savings requires further workflow-specific validation.
\end{revision_1}

Neither volumetric nor surface-based geometric metrics are designed to reliably predict the dosimetric consequences of segmentation deviations. 
A marginal contour expansion or contraction can alter dose-volume histogram (DVH) parameters and organ-specific dose constraints, potentially affecting clinical assessments of toxicity risk and target coverage, without any corresponding shift in DSC or surface distance. 
A recent large-scale benchmarking analysis from the HaN-Seg challenge \cite{podobnik2024han} explicitly confirmed this disconnect, finding no clear correlation between geometric accuracy and dosimetric compliance across submitted deep-learning algorithms \cite{podobnik2026dosimetric}. 
As the field progresses toward foundation models and agentic pipelines whose outputs will directly affect treatment planning workflows, embedding dosimetric assessment into standard benchmarking can be recommended as a prerequisite rather than an optional supplement of model evaluation.

\begin{revision_1}
Going beyond geometric and dosimetric agreement, readiness for clinical use depends on endpoints that are rarely reported for promptable and vision--language models including physician correction time ~\cite{lustberg2018clinical} and blinded expert acceptability, reduction of interobserver variability, feasibility within the time constraints of online adaptive workflows~\cite{kalinauskaite2025online}. 
In addition, external multi-institutional validation and prospective evaluation in clinical workflows are needed to establish generalizability and practical utility~\cite{rong2024nrg}.
So far, the models discussed in Sections~\ref{sec:promptable}--\ref{sec:agentic} have mostly been evaluated retrospectively, often on single-institution data and with geometric metrics alone, which limits conclusions about their clinical utility.
\end{revision_1}

\subsection{Beyond Dice: Target Volume Embeddings and Outcome Analysis}
A complementary direction is to compare a predicted contour with its reference in feature space rather than relying solely on geometric overlap. 
Such features can be imaging biomarkers, such as radiomics-derived features \cite{lambin2012radiomics}, or learned latent representations of neural networks. 
For instance, imaging biomarkers can be extracted from the same image volume twice, once under the predicted mask and once under the reference mask. 
The divergence between the two resulting feature vectors then quantifies the impact of the segmentation error on these biomarkers \cite{konz2026frechet}.

Going beyond Dice, such divergences can quantify the \textit{semantic and downstream impact} of a segmentation difference. 
Geometric metrics treat all surface deviations symmetrically, yet over-delineation (including surrounding healthy tissue) and under-delineation (cropping into target tissue) produce asymmetric perturbations to intensity, texture, and shape features, and therefore result in asymmetric consequences for any downstream analysis that consumes those features. 
The same evaluation logic can be applied to the richer multimodal embedding spaces learned by VLMs, offering a representation that encodes not only geometric properties but also semantic textual context, potentially allowing improved interpretation of clinically consequential contour differences based on extracted feature vectors.

Beyond contour comparison, embedding-space representations naturally support similarity-based retrieval, in which the nearest neighbors of a query case can be identified to select comparable patients from a historical cohort, for instance, to inform clinical decision-making through prior treatment outcomes.
Recent work on multimodal temporal patient embeddings has demonstrated such retrieval at healthcare-system scale \cite{zhang2026multimodal}, suggesting that analogous approaches applied to radiotherapy-specific VLM embeddings are technically feasible.

\begin{figure}[htbp]
  \centering
  \includegraphics[width=\textwidth]{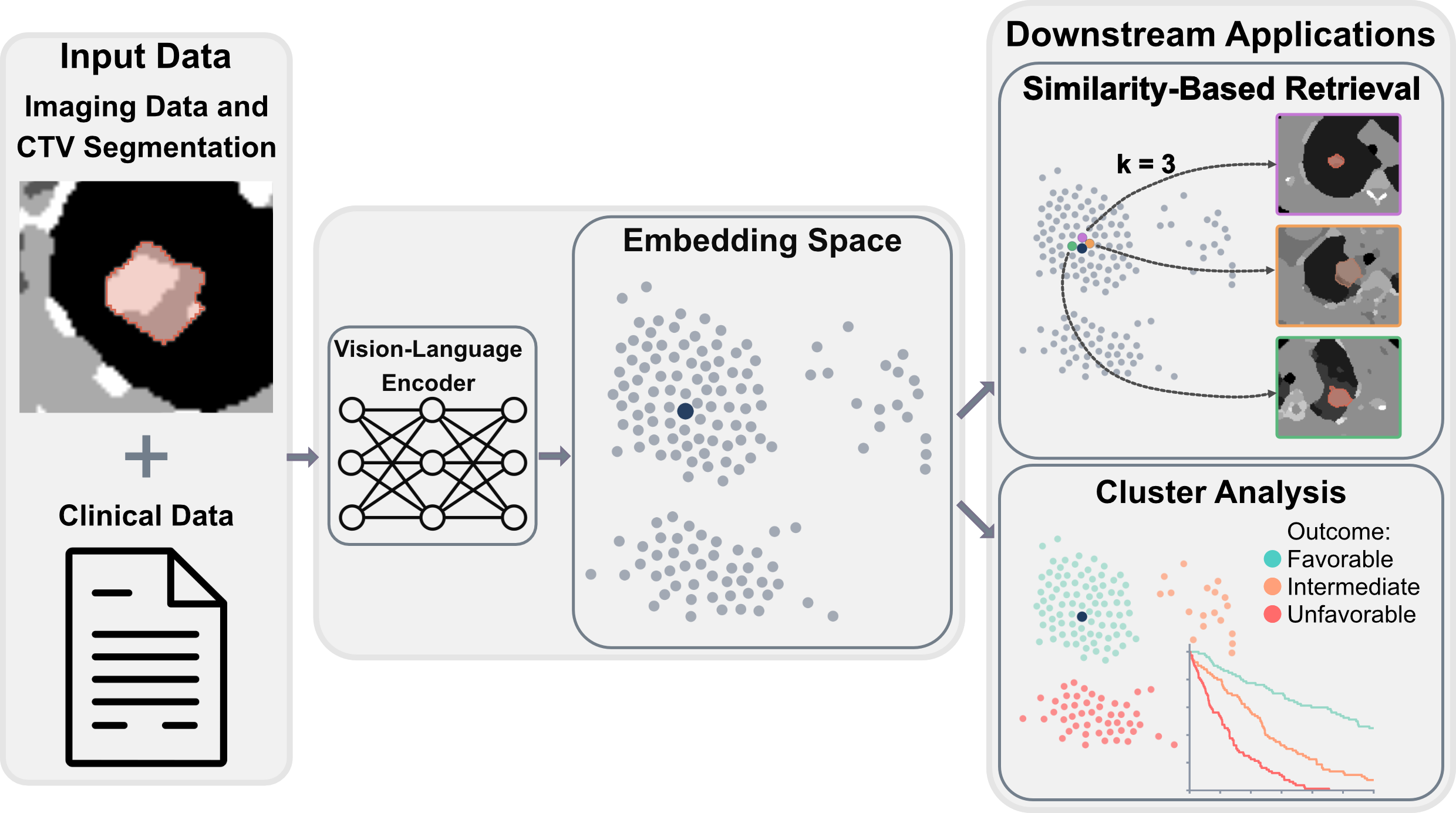}
  \caption[Multimodal target volume embeddings and downstream evaluation.]{\textbf{Multimodal target volume embeddings and downstream evaluation.} 
  Imaging data and clinical metadata are jointly encoded into a shared embedding space via a vision--language encoder. 
  The resulting representations enable downstream applications such as similarity-based retrieval of analogous cases and cluster analysis linked to patient outcomes.}
  \label{fig:embedding}
\end{figure}

Taken further, patient-level outcomes could serve as the ultimate benchmark for segmentation quality, rather than contour agreement or dosimetric compliance. 
Applying this to VLMs introduces a qualitatively new possibility: the joint embedding space of a VLM trained on large-scale annotated radiotherapy data can learn to compress morphological, semantic, and contextual information about target volumes into a representation that \begin{revision_1}we hypothesize to be \end{revision_1} prognostically richer than any single descriptor.
\begin{revision_1}This hypothesis could be tested by linking VLM contour embeddings to outcome information to examine whether they stratify patients by outcome and, ultimately, which target volume definitions are outcome-optimal (Figure~\ref{fig:embedding}). For any such association, however, one would need to assess the extent to which they are confounded by treatment and patient selection.
\end{revision_1}

A longer-horizon extension would involve jointly training the model on segmentation, temporal alterations in treatment volume, as well as clinical outcomes, thereby actively encoding treatment-predictive information in the latent space that can subsequently be localized through targeted probing or counterfactual interventions. 
Such an approach requires linked longitudinal imaging and outcome datasets of a scale not yet routinely available in public radiotherapy data, and it further demands controls for treatment and patient-selection confounding. 
If realized, such an approach could help answer not only whether a predicted contour matches the reference, but also whether patients delineated with that contour fare better. 
Ultimately, target volumes could then be optimized on the basis of outcome evidence rather than anatomical convention alone.

\subsection{Emerging Measures: From Prompt Impact To Agent Trajectory Analysis} \label{sec:agent_eval}

Due to the recency of foundation VLMs, it is an open question for benchmarking standardization to reproducibly establish verification and quantification of text prompt influence on a model's output.
When a segmentation VLM is evaluated by comparing its predicted mask to a reference mask using DSC, a perfect score is achievable even if the model entirely ignores the text and relies exclusively on strong visual priors.
Studies in promptable segmentation have demonstrated that models can be highly sensitive to semantically trivial prompt variations, where minor perturbations to bounding box coordinates alter \textit{SAM}-like model outputs substantially and unpredictably \cite{moskalenko2026breps}, and semantically equivalent text reformulations of the same concept produce inconsistent masks in text-guided segmentation \cite{wu2026promptgroup}. 

Testing prompt adherence in radiotherapy, the \textit{LLMSeg} authors modified the clinical input to confirm that the corresponding output changed in a clinically expected direction \cite{oh2024llm}, which is a promising approach but needs to be further developed, motivating the following recommendation of a systematic evaluation paradigm.  

\textit{Counterfactual prompt evaluation} systematically applies contrastive prompt pairs to the same image with opposed clinical attributes that should change the target volume, such as altered nodal staging, switching laterality, or adding high-risk subvolumes to verify that the predicted mask changes in the expected direction and magnitude. 
Further \textit{language ablation} tests can compare outputs produced with and without the text condition across a held-out test set, quantifying how much of a model's performance is attributable to the prompt versus the image alone. 
Additionally, \textit{semantic invariance testing} requires that paraphrased prompts referencing the same clinical concept produce geometrically consistent segmentations.

These approaches collectively address what can be framed as an alignment problem \cite{wiener1960some,amodei2016concrete} defined by the challenge of ensuring that a model acts on the user's actual intent encoded in the prompt rather than on proxy patterns that its training rewarded. 
In the radiotherapy context, such an alignment gap has direct clinical stakes: a model that produces anatomically plausible CTV contours for the wrong clinical reason can result in systematically incorrect volumes for patients who deviate from the training distribution, precisely the cases in which correct language grounding matters most. 
This highlights the need for standardized benchmarks for prompt adherence in 3D radiotherapy segmentation, while accounting for the geometric and dosimetric evaluations discussed above.

These evaluation challenges are further compounded when the segmentation model is embedded within an agentic pipeline, which, rather than producing a single input-output pair, unfolds a multi-step trajectory across tool calls, retrieval steps, and potential interactions with human expert reviewers. 
This represents a qualitatively different evaluation problem for which a clinical standard, particularly in radiotherapy, has yet to be established \cite{mehandru2023agents}.
For instance, a final agent-created geometrically acceptable contour provides no information about whether the agent retrieved the correct guideline, whether the retrieved passage was correctly interpreted, or whether the segmentation model acted on the right instructions, as errors at any step can cancel out or compound unpredictably.

We thus recommend a \textit{trajectory-level evaluation} that assesses the correctness of intermediate steps alongside measures of \textit{graceful degradation}, i.e., whether the system flags uncertainty and defers safely to the clinical expert when it encounters out-of-distribution inputs rather than producing a plausible but incorrect contour.
Emerging medical agent benchmarks, such as MedAgentBoard \cite{zhu2025medagentboard}, have initiated the establishment of systematic frameworks to this end, finding that existing evaluations poorly cover real-world clinical workflow conditions. 
This indicates a need for prospective, clinically meaningful simulation studies that include edge cases and guideline ambiguities to define potential safety claims for agentic radiotherapy planning.

\section{Discussion and Conclusions}

The past two decades have witnessed significant progress in segmentation methodology for radiotherapy, from the early atlas-based approaches that first demonstrated the feasibility of automatic delineation, through the deep learning revolution \begin{revision_1} sparked and driven by deep convolutional networks \cite{krizhevsky2012imagenet} and \textit{U-Net} \cite{ronneberger2015u} \end{revision_1} architectures, to the emerging paradigm of \begin{revision_1}transformer-based \cite{vaswani2017attention}\end{revision_1} promptable, language-conditioned foundation models, and, \begin{revision_1}most recently, first concepts for agent-based systems.\end{revision_1}
Each methodological shift solved some problems while introducing new ones: atlas-based methods encoded anatomy but were brittle to variation, task-specific \textit{CNNs} improved performance  \begin{revision_1} and continue to be strong baselines on traditional geometric metrics such as DSC\end{revision_1}, but remained narrow \begin{revision_1}and without training-free portability to any new domain, modality or organ. 
The subsequent,\end{revision_1} newer promptable models increase flexibility, \begin{revision_1}and versatility through zero-shot generalization\end{revision_1}, but are not yet \begin{revision_1}found to be validated, \end{revision_1}robust \begin{revision_1}and prompt-controllable \end{revision_1}enough for routine clinical use.

Nevertheless, a critical gap persists between research progress and clinical reality. 
In routine radiotherapy workflows, clinically deployed auto-segmentation tools still appear to be dominated by atlas-based systems and commercial task-specific deep-learning auto-contouring packages \cite{doolan2023clinical, heilemann2023clinical, robert2021clinical, zabel2021clinical}.
\begin{revision_1}Despite their narrow application scope, \end{revision_1} they are well-validated, regulatory-approved, and integrated into clinical workflows.
A recent survey of radiotherapy departments in German-speaking countries found that 65.7\% already use AI-based auto-contouring software in clinical routine \cite{vorbach2025contouring}.

The more recent advances discussed in this review, such as promptable foundation models, VLMs, multimodal approaches such as \textit{LLMSeg}, and the agentic architectures outlined in Section \ref{sec:agents}, remain largely at the proof-of-concept or early research stage\begin{revision_1}, although this may change\end{revision_1}. 
\begin{revision_1} At the present point in time,\end{revision_1} their generalizability across institutions, imaging protocols, and patient populations has not yet been established at the scale required for routine clinical deployment, and prospective clinical validation data are scarce. Moreover, regulatory frameworks for such foundation model-based systems are unclear, as not every input used for training a foundation model can realistically be validated in the same manner as for task-specific deep learning models.

A related challenge concerns post-deployment adaptation. 
In clinical practice, contouring guidelines are updated, new imaging hardware is introduced, and institutional protocols evolve. 
Agentic systems are inherently designed to accommodate such changes without requiring retraining of the underlying models, \begin{revision_1} a key advantage over static task-specific deep-learning auto-contouring approaches\end{revision_1}. 
However, under current regulatory frameworks such as CE certification, any modification to a certified system may necessitate a new conformity assessment, creating a tension between the dynamic adaptability of agentic architectures and the static nature of regulatory approval. 
Defining reproducible evaluation protocols for systems whose outputs depend on dynamically retrieved external knowledge, as discussed in Section \ref{sec:agent_eval}, is a prerequisite for resolving this regulatory tension.

The direction of travel is clear, but the gap between methodological flexibility and clinically reliable deployment remains substantial.
Multimodal foundation models capable of integrating imaging, clinical text, and guideline knowledge represent a possible path toward segmentation systems that not only automate contouring but actively support the \begin{revision_1}interpretability, versatility, and\end{revision_1} clinical reasoning underlying contour definition.
This could include communicating the rationale for proposed contours in domain-specific language, incorporating tumor stage and treatment history into target volume decisions, and transparently conveying the uncertainties associated with each output.
Realizing this vision, however, will require evaluation frameworks that match the complexity of the systems they assess. 
Geometric overlap metrics alone cannot capture whether a model acts on the clinical reasoning encoded in its prompt, whether an agentic pipeline retrieves and applies the correct guideline, or whether the resulting contour is dosimetrically adequate. 
Advancing segmentation methods \begin{revision_1} alongside their evaluation methods\end{revision_1} will be essential for translating these capabilities into clinical practice \begin{revision_1} and realizing the benefits of next-generation interactive, agentic, multimodal radiotherapy workflows.\end{revision_1}

\subsection*{Declaration of Generative AI and AI-assisted Technologies 
in the Writing Process}

During the preparation of this work, the authors used Claude (Anthropic) for language editing, literature search, and figure caption drafting, and Gemini (Google) via PaperBanana \cite{zhu2026paperbanana} for figure drafting.
After using these tools, the authors reviewed and edited the content as needed and take full responsibility for the content of the publication.


\bibliographystyle{elsarticle-num} 
\bibliography{references}

\end{document}